\documentclass{edp-conf}

\usepackage{graphics}
\usepackage[OT1]{fontenc}

\newcommand{\noi}{ \noindent }

\newcommand{\beq}{\begin{equation}}
\newcommand{\eeq}{\end{equation}}

\newcommand{\greq}{\begin{equation}\left\{ \begin{array}{l}}
\newcommand{\egreq}{\end{array}\right. \end{equation}}

\newcommand{\YL}{ Y^m_\ell }

\newcommand{\eeqn}[1]{\label{#1}\end{equation}}

\newcommand{\beqa}{\begin{eqnarray*}}
\newcommand{\eeqa}{\end{eqnarray*}}

\begin{document}
 
\TitreGlobal{SF2A 2001} 
 
\title{Acoustic modes in spheroidal cavities}

\author{F. Ligni\`eres}
\address{
Laboratoire d'Astrophysique, CNRS UMR 5572, Obs. Midi-Pyr\'en\'ees,
57 avenue d'Azereix, 65008 Tarbes Cedex, France}
\author{M. Rieutord}
\address{
Laboratoire d'Astrophysique de Toulouse,
Observatoire Midi-Pyr\'en\'ees, 14 avenue E. Belin, 31400 Toulouse,
France and Institut Universitaire de France}
\author{L. Valdettaro}
\address{
Dipartimento di Matematica, Politecnico di Milano, Piazza L.
da Vinci, 32, 20133 Milano, Italy
}

\runningtitle{Acoustic modes in spheroidal cavities}

\maketitle

\begin{abstract}
Oscillation modes of rapidly rotating stars have not yet been calculated
with precision, rotational effects being generally approximated by
perturbation methods. We are developing a numerical method able to
account for the deformation of the star by the centrifugal force and,
as a first step, we determined the acoustic modes of a uniform density
spheroid and studied how the spheroid flatness affects these modes.
\end{abstract} 

\section{Introduction}

The ratio of centrifugal and gravity forces being generally small in
stars, the effect of rotation on gravito-acoustic stellar oscillations
has been mostly studied with perturbation methods.  Although fully
justified in the context of helioseismology, this approach might not be
accurate enough for rapidly rotating stars when the difference between the
correct and approximated frequency becomes comparable with the accuracy
of frequency measurements.  We thus expect that new non-perturbative
approaches will be necessary in the context of future asteroseismology
missions (Mons, Corot, Eddington).

Such non-perturbative calculations should include the effects of the
Coriolis and centrifugal forces on the wave motions as well as the
effect of the centrifugal force on the equilibrium state of the star.
This latter effect breaks down the spherical symmetry of the cavity
within which acoustic modes resonate and this introduces mathematical
difficulties since the related eigenvalue problem is no longer fully
separable (except in the particular case of spheroids).

In this paper, we present a numerical method able to solve this type of
problem. Before it is applied to a realistic stellar model, we tested
it in a simpler case. We choose the homogeneous spheroid because an
alternative method exists in this case. Let also recall that stationary
solutions of a self-graviting and uniformly rotating gas of constant
density are spheroids (the so-called Maclaurin spheroids).  Thus acoustic
modes of homogeneous spheroids can be viewed as the oscillation modes
of Maclaurin spheroids in the high frequency limit (in order to neglect
the effect of Coriolis force on the oscillatory motions) and within the
Cowling approximation (to neglect the effect of gravity on the modes).
In the next sections, the formalism is presented and the result of the
test as well as a first analysis of the flatness effects on the frequency
spectrum are given.

\section{The formalism}

Oscillatory motions of small amplitude in a perfect fluid of uniform
density are governed by the Helmholtz equation,

\begin{equation}
\Delta \hat{\psi} + k^2 \hat{\psi} = 0,
\label{eqdl2} 
\end{equation}
\noindent where $\psi = \hat{\psi} \exp(i \omega t)$ is the velocity
potential (${\bf u} = \nabla \psi$) and $k = \omega/c_s$, the ratio of
the mode frequency, $\omega$, and the sound speed, $c_s$.  The boundary
condition at the surface of the cavity is, ${\bf n}\cdot\nabla \hat{\psi}
= 0,$ where ${\bf n}$ is a vector normal to the surface.

Expect for the special cases of spherical or spheroidal cavities this
eigenvalue problem is not fully separable.  Then, for arbitrary axially
symmetric surfaces, a 2D eigenvalue problem must be solved numerically.
We first note that, for numerical reasons, it is preferable to apply
the boundary conditions on a coordinate surface.  A coordinate
system must thus  be chosen so that the surface be a surface of
coordinate ($\zeta = cste$).  That is, if $r= S(\theta)$ describes
the surface, suitable coordinates are $(\zeta = f(\zeta, \theta),
\theta, \phi)$ where for some value  $\zeta_0$ the function $f$ verifies
$r=f(\zeta_0,\theta)=S(\theta)$.  In this expressions $(r, \theta, \phi)$
are the usual spherical coordinates.  Following Bonazzola et al. (1998),
$f$ is then specified so that the regularity conditions at the center
have a simple form.  For spheroids, we used,

\begin{equation}
r = f(\zeta, \theta) = R_e \left[\zeta + (3 \zeta^4 - 2 \zeta^6)\left(\frac{1}{\sqrt{1 + g(\epsilon) \cos^2\!\theta}} - 1\right)
 \right]
\end{equation}
\noindent 
where $R_e$ and $R_p$ are respectively the semi-major axis and the
semi-minor axis, $\epsilon = 1 - R_p / R_e$ is the flatness and
$g(\epsilon) = \epsilon(2-\epsilon)/(1-\epsilon)^2$.  The eigenvalue
problem is written with these new coordinates.  By expanding the
solution $\hat{\psi}(\zeta,\theta,\phi)$ on spherical harmonics and
projecting the equations on each spherical harmonic $\YL(\theta,\phi)$,
one obtains for each value of the azimuthal number $m$ two sets of ODE
of the variable $\zeta$ coupling the coefficients of spherical harmonic
expansion  corresponding respectively to odd and even degree numbers,
that is ($\hat{\psi}^{m+2k}_m (\zeta), \; 0 \leq k < +\infty$)  and
$(\hat{\psi}^{m+2k+1}_m(\zeta),  \; 0 \leq k < +\infty)$.  Then following
a method developed by Rieutord and Valdettaro (1997) after Orszag (1971),
the ODE are discretized on the Gauss-Lobato grid associated with Chebyshev
polynomials. A set of linear equations dependent of the parameter $k$
results and non-trivial solutions of this system are found using either
the QZ algorithm or the Arnoldi-Chebyshev algorithm.

To test this general method, we considered a spheroidal cavity as a simple
alternative method exists in this case.  This is because  the eigenvalue
problem is separable using the so-called oblate spheroidal coordinates
$(\xi, \eta, \phi)$ defined as $(x = a\cosh\xi\sin\eta\sin\phi, \; y =
a\cosh\xi\sin\eta\cos\phi, \; z = a \sinh\xi\cos\eta)$, where $0\leq \xi
<+ \infty$, $0\leq \eta \leq \pi$ et $0\leq \phi \leq 2 \pi$.  The package
Linear Solver Builder developed by Rieutord and Valdettaro was also used
to solve the eigenvalue problem written in this coordinate system.

Finally, we tested our results in the range of small flatness using
a perturbation approach.  Following the method described in Gough et
al. (1990), we calculated analytically the first order effect of the
flatness on the mode frequencies. It reads,

\begin{equation}
\tilde{\omega}_{n\ell m} = \tilde{\omega}^0_{n\ell} + 
\frac{2 \epsilon \tilde{\omega}^0_{n\ell}}{3 (2\ell -1)(2\ell +3)}\left( 
\ell(\ell +1) - 3 m^2 + 3 \frac{\ell(\ell +1) -3m^2}
{ {\left({\tilde{\omega}}^{0}_{n\ell}\right)}^2 - \ell(\ell +1)} \right),
\label{eqdl} 
\end{equation}
\noi where $\tilde{\omega}^0_{n\ell} = {\omega}^0_{n\ell} R/ c_s$
is the dimensionless eigenfrequency corresponding to the sphere of the
same volume, $R = R_e \left( 1 - \epsilon \right)^{1/3}$. It is given as
the $n^{\mbox{th}}$ root (by growing order) of the following equation,
$(\ell+1) J_{\ell+1/2}(x) - x J_{\ell-1/2}(x) = 0, \; 1 \leq \ell <
+\infty$, where $J_{\ell}$ are Bessel functions.

\section{Results}

We found that the general method and the method specific
to spheroidal cavities provide the same frequencies with
a high level of accuracy for arbitrary value of the flatness
between $0$ and $0.5$.
Moreover, as the flatness goes to zero, the frequencies 
converge towards the values given by the asymptotic analysis (Eq. \ref{eqdl}).
We are thus encouraged in using the general method described in this paper
for future pulsation
models which shall take
into account the deformation of the star equilibrium
by the centrifugal force.

We also studied 
how the flatness of the spherical cavity affect the 
frequency of acoustic modes.
The first point is to understand why, for increased flatness,
the frequency of some
modes increases while the frequency of others decreases.
Simple arguments concerning the region of mode propagation
can explain these behaviors.
For large degree ($\ell \gg 1$) quantitative results can even
be deduced from such arguments: Large degree axi-symmetric modes ($\ell \gg 1,
m=0$) propagate in meridional planes just below  the cavity surface.
Therefore, the characteristic length
involved in their quantization 
condition shall be the perimeter of the circle, $2 \pi R$, for the sphere,
and the perimeter of the ellipse
for the spheroid. 
By contrast sectorial modes ($m=\ell$) are confined towards
the equator so that the characteristic length is the perimeter
of the equatorial circle that is $2 \pi R$ for the sphere and $2\pi R_e$
for the spheroid.
The increase or decrease of the modes frequency should
then be determined by the ratio of these
characteristic lengths.
This interpretation is consistent with the first order asymptotic formula 
(Eq. \ref{eqdl}) since in the large $\ell$ limit, the frequency ratio
$\tilde{\omega}_{n\ell m} / \tilde{\omega}^0_{n\ell}$
equals the ratio of the characteristic lengths.

As a consequence of the differential effect of flatness
on frequencies, modes with originally distinct frequency
tends to reach identical frequencies when flatness increases. 
However, this is not
allowed for the coupled modes and avoided crossing occur.
As all the modes having identical azimuthal number
and degree numbers of the same parity are coupled (see above),
avoided crossings occur at a strong rate in such a way
that 
for non negligible 
flatness most of the modes have
experienced many avoided crossing.

Another interest of the present calculation
is to assess the limit of validity of the perturbative approach.
Note first that, for small rotation rates, the flatness of MacLaurin spheroids
is proportional to the ratio of centrifugal and gravity forces.
Thus first order effects in terms of flatness correspond to second order effect
in terms of the rotation rate.
We found that the difference between the correct and the approximated frequency
depends on the type of modes. But it can be relatively important.
For example, at $\epsilon = 0.1$, the relative difference
for axisymmetric modes of degree $\ell = 1$ grows with
the radial order and for $1 \leq n \leq 6$ it is comprised between
$0.01$ and $0.03$.
If we extrapolate such relative differences to a rapidly rotating $\delta$ Scuti
oscillating around 200 $\mu$ Hz, we find a frequency difference of 1 to 3 $\mu$ Hz that
is one order of magnitude larger than the accuracy planned for
Corot mission.
Although a firm conclusion should await stellar oscillation 
models, the present estimate points towards the necessity of 
using non-perturbative method for spatial seismology of
rapidly rotating stars.

We also noticed the apparition of new types of modes
in the spheroidal geometry.
They are easily identified
in the context of ray theory of axi-symmetric modes.
Whereas for spherical cavity, all caustics are concentric spheres,
two type of caustics, concentric spheroids and hyperboloids,
are possible within a spheroidal cavity.
In this last case, the ray does not propagate in the equatorial regions
comprised between the surface boundary and the hyperbolic caustic.
With our calculation we effectively found modes which show
hyperbolic like regions of very low amplitude around the equator.

\section{Future work}

The next step is to calculate the oscillations
of a rotating polytropic star including the effect of Coriolis 
and centrifugal forces on the motions. Among the new features
arising when considering these more realistic stellar models
is the chaos of rays 
dynamics within non-spherical and non-spheroidal cavities. 
Astrophysical consequences of this 
chaotic behavior shall be investigated.

\end{document}